\documentclass[aps,prd,groupedaddress,showpacs,nofootinbib]{revtex4}
\usepackage{graphicx,amssymb}
\usepackage{epsfig}
\usepackage{amsmath}
\usepackage{amsfonts}
\usepackage{amssymb}
\usepackage{graphicx}
\usepackage{colordvi}
\begin{document}
\title{ Spherical symmetry in  $f(R)$-gravity}
\author{S. Capozziello\footnote{e\,-\,mail address:
capozziello@na.infn.it}$^{\diamond}$,\,A. Stabile\footnote{e -
mail address: stabile@sa.infn.it}$^{\natural}$, \,A.
Troisi\footnote{e\,-\,mail address:
antrois@gmail.com}$^{\diamond}$}

\affiliation{$^{\diamond}$ Dipartimento di Scienze Fisiche and
INFN, Sez. di Napoli, Universit\`a di Napoli "Federico II",
\\
Compl. Univ. di Monte S. Angelo,
\\Edificio
G, Via Cinthia, I-80126 - Napoli, Italy }

\affiliation{$^{\natural}$ Dipartimento di Fisica "E. R.
Caianiello", Universita' degli Studi di Salerno,\\
 Via S. Allende, I-84081 Baronissi (SA), Italy.}

\begin{abstract}
Spherical symmetry in $f(R)$ gravity is discussed in details
considering also  the relations with the weak field limit. Exact
solutions are obtained for constant Ricci curvature scalar and for
Ricci scalar depending on the radial coordinate. In particular, we
discuss how to obtain results which can be consistently compared
with General Relativity giving the well known post-Newtonian and
post-Minkowskian limits. Furthermore, we implement a perturbation
approach to obtain solutions up to the first order starting from
spherically symmetric backgrounds. Exact solutions are given for
several classes of $f(R)$ theories in both $R\, =$ constant and
$R\,=\,R(r)$.
\end{abstract}
 \maketitle

\section{Introduction}

The recent advent of  new cosmological precision tests, capable of
probing physics at very large redshifts, has changed the modern
view of cosmos  ruling out the Cosmological Standard Model based
on General Relativity (GR), radiation and baryonic  matter. Beside
the introduction of dark matter, needed to fit the astrophysical
dynamics at galactic and cluster scales (i.e. to explain clustered
structures), a new ingredient is requested  in order to explain
the observed accelerated behavior of the Hubble flow: the dark
energy.

In particular, the luminosity distance of Ia Type Supernovae
\cite{sneIa}, the Large Scale Structure \cite{lss} and the
anisotropy of Cosmic Microwave Background \cite{cmbr} suggest that
the widely accepted Cosmological Concordance Model ($\Lambda$CDM)
is a spatially flat Universe,  dominated by cold dark matter (CDM
$(\sim 0.25\div 0.3\%)$ which should explain the clustered
structures) and dark energy $(\Lambda$ $\sim 0.65\div 0.7\%)$, in
the form of  an ``effective" cosmological constant, giving rise to
the accelerated behavior.

Although the cosmological constant  \cite{padma2003,PR03,copeland}
remains the most relevant candidate to interpret the  accelerated
behavior, several proposals have been suggested in the last few
years: quintessence models, where the cosmic acceleration is
generated by means of a scalar field, in a way similar to the
early time inflation, acting at large scales and recent epochs
\cite{quintessence}; models based on exotic fluids like the
Chaplygin-gas \cite{chaplygin}, or non-perfect fluids
\cite{cresc}); phantom fields, based on scalar fields with
anomalous signature in the kinetic term \cite{phantom}, higher
dimensional scenarios (braneworld) \cite{brane}. Actually, all of
these models, are based on the peculiar characteristic of
introducing  new sources into the cosmological dynamics, while it
would be preferable to develop scenarios  consistent with
observations without invoking further parameters or  components
non-testable (up to now)  at a fundamental level.

The resort to modified gravity theories, which extend in some way
the GR, allows to pursue this different approach (no further
unknown sources) giving rise to suitable cosmological models where
a late time accelerated expansion naturally arises.

The idea that the Einstein gravity should be extended or corrected
at large scales (infrared limit) or at high energies (ultraviolet
limit) is suggested by several theoretical and observational
issues. Quantum field theories in curved spacetimes, as well as
the low energy limit of string theory, both imply
semi\,-\,classical effective Lagrangians containing  higher-order
curvature invariants or scalar-tensor terms. In addition, GR has
been tested only at  solar system scales while it shows several
shortcomings if checked at higher energies or larger scales.

Of course modifying the gravitational action asks for several
fundamental challenges. These models can exhibit instabilities
\cite{instabilities-f(R)} or ghost\,-\,like behaviors
\cite{ghost-f(R)}, while, on the other side, they should be
matched with the low energy limit observations and experiments
(solar system tests, PPN limit). Despite of all these issues, in
the last years, several interesting results have been achieved in
the framework of the so called $f(R)$- gravity  at cosmological,
galactic and solar system scales.

For example, models based on generic functions of the Ricci scalar
$R$ show cosmological solution with late time accelerating
dynamics \cite{f(R)-cosmo}, in addition,  it has been shown that
some of them could agree with CMBR observational prescriptions
\cite{barrow07}, nevertheless this matter is still argument of
debate \cite{bean-f(R)}. For a review of the models and their
cosmological applications see, e.g.,\cite{GRGrev,odirev}.

Moreover, considering $f(R)$-gravity in the low energy limit, it
is possible to obtain corrected gravitational potentials capable
of explaining the flat rotation curves of spiral galaxies without
considering huge amounts of dark matter
\cite{noi-mnras,salucci,sobouti,mendoza} and, furthermore, this
seems the only self-consistent way to reproduce the universal
rotation curve of spiral galaxies \cite{salucci2}. On the other
hand, several anomalies in solar system experiments could be
framed and addressed in this picture \cite{bertolami}.

In the last few years, several authors have dealt with this matter
considering  the Parameterized Post Newtonian (PPN) limit
\cite{ppn-tot}. On the other hand, the investigation of
spherically symmetric solutions for such kind of models has been
developed in several papers
\cite{spher-f(R),multamaki06,kainulainen,noether}. Such an
analysis deserves particular attention since it can allow to draw
interesting conclusions on the effective modification of the
gravitational potential induced by higher order gravity at low
energies and, in addition, it could shed new light on the PPN
limit of such theories. However, in several recent papers, it has
been shown that several classes of $f(R)$ models fairly evade the
Solar System constraints and agree with the limits of
PPN-parameters, in particular with $\gamma\sim 1$ probed by
experiments \cite{Hu,Star,Appleby,Tsuji,Odintsov}. This is not the
argument of the present paper, however it is important to stress
that the spherical symmetry and the weak filed limit have to be
carefully considered in order to find out physically viable
models.

In this paper, we are going to analyze, in a general way and
without specifying {\it a priori} the form of the Lagrangian, the
relation between the spherical symmetry and the weak field limit
of $f(R)$ gravity. Our aim is to develop a systematic approach
considering the theoretical prescriptions to obtain a correct weak
field limit in order to point out  the analogies and the
differences with respect to GR. A fundamental issue is to recover
the asymptotically flat solution in absence of gravity and the
well-known results related to the specific case $f(R)=R$, i.e. GR.
Only in this situation a correct comparison between GR and any
extended gravity theory is possible from an experimental and a
theoretical  viewpoint. For example, as already shown in
\cite{Newf(R)} and discussed in \cite{clifton}, the Birkhoff
theorem is not a general result for fourth-order models also if it
holds for several interesting classes of these theories as
discussed, for example, in \cite{noether,whitt,barraco}.

The layout of the paper is the following. Sec.II is devoted to
some general remarks on spherical symmetry in $f(R)$ gravity. In
particular, we derive the field equations in such a symmetry.  In
Sec.III,  the expression of the Ricci scalar and the general form
of metric components are derived in spherical symmetry discussing
how recovering the correct Minkowski flat limit. Sec.IV is devoted
to the discussion of spherically symmetric background solutions
with constant scalar curvature considering, in particular
Schwarzschild-like and Schwarzschild-de Sitter-like solutions with
constant curvature. Some remarkable $f(R)$ models are worked out
as example. In Sec. V, we discuss the cases in which the spherical
symmetry is present also for the Ricci scalar depending on the
radial coordinate $r$. This is an interesting situation, not
present in GR. In fact, as it is well known, in the Einstein
theory, the Birkhoff theorem states that a spherically symmetric
solution is always stationary and static \cite{ellis} and  the
Ricci scalar is constant. In $f(R)$ the situation is more general
and then the Ricci scalar, in principle, can evolve with radial
and time coordinates. Sec.VI is devoted to the study of a
perturbation approach starting from a spherically symmetric
background considering the general case in which the Ricci scalar
is $R=R(r)$. The motivation is due to the fact that, in GR, the
Schwarzschild solution and the weak field limit coincide under
suitable conditions. This could be a good test bed to construct a
well-posed weak field limit for any extended theory of gravity.
Finally, we work out some $f(R)$-models finding
spherically-symmetric solutions by the perturbation approach.
Conclusions are drawn in Sec.VII.

\section{Spherical symmetry}

Let us consider an analytic function $f(R)$ of the Ricci scalar
$R$.  The variational principle for this action is:

\begin{equation}\label{fRaction}
\delta\int
d^4x\sqrt{-g}\biggl[f(R)+\mathcal{X}\mathcal{L}_m\biggr]\,=\,0
\end{equation}

where ${\displaystyle \mathcal{X}=\frac{16\pi G}{c^4}}$,
$\mathcal{L}_m$ is the standard matter Lagrangian and $g$ is the
determinant of the metric. By varying  with respect to the metric,
we obtain the field equations \footnote{It is possible to take
into account also the Palatini approach in which  the metric $g$
and the connection $\Gamma$  are considered independent variables
(see for example \cite{palatini}). Here we will consider the
Levi-Civita connection and will use the metric approach. See
\cite{GRGrev,ACCF} for a detailed comparison between the two
pictures.}

\begin{equation}\label{HOEQ}
f'(R)R_{\mu\nu}-\frac{1}{2}f(R)g_{\mu\nu}-f'(R)_{;\mu\nu}+g_{\mu\nu}\Box
f'(R)\,=\,\frac{\mathcal{X}}{2}T_{\mu\nu}\,,
\end{equation}
where the trace equation read
\begin{equation}\label{TrHOEQ}
3\Box f'(R)+f'(R)R-2f(R)\,=\,\frac{\mathcal{X}}{2}T\,,
\end{equation}
and
$T_{\mu\nu}\,=\,\displaystyle\frac{-2}{\sqrt{-g}}\frac{\delta(\sqrt{-g}\mathcal{L}_m)}{\delta
g^{\mu\nu}}$ is the energy momentum tensor of standard matter,
$T={T^\sigma}_\sigma$ is the related trace and
$f'(R)=\frac{df(R)}{dR}$.  Eqs.(\ref{HOEQ}) can be rewritten in an
Einstein-like form  recasting the higher than second order
contributions under the form of an effective stress\,-\,energy
tensor of geometrical origin:

\begin{equation} \label{eq: field}
G_{\mu \nu} = R_{\mu \nu} - \frac{1}{2} R g_{\mu \nu} =
T^{(curv)}_{\mu \nu} + T^{(m)}_{\mu \nu}\,.
\end{equation}
The effective curvature  stress\,-\,energy tensor is
\cite{f(R)-noi}\,:

\begin{equation}\label{eq: curvstress}
T^{(curv)}_{\mu\nu} = \frac{1}{f'(R)} \left\{ g_{\mu\nu}
\left[f(R) - R f'(R)\right] + f'(R)^{; \rho \sigma}
\left(g_{\mu\rho} g_{\nu\sigma} - g_{\rho\sigma} g_{\mu
\nu}\right)\right\}\,.
\end{equation}
The matter term enters Eqs.(\ref{eq: field}) through the modified
stress\,-\,energy tensor\,:

\begin{equation}\label{eq:
mattstress} T^{(m)}_{\mu\nu} =
\frac{\mathcal{X}T_{\mu\nu}}{2f'(R)}\,,
\end{equation}
where there is a non-minimal coupling between matter and geometric
degrees of freedom. For the purposes of this paper, we will refer
to the field equations in the form (\ref{HOEQ}) which reveal to be
more useful for the following considerations.

The most general spherically symmetric metric can be written as
follows\,:

\begin{equation}\label{me0}
ds^2\,=\,m_1(t',r')dt'^2+m_2(t',r')dr'^2+m_3(t',r')dt'dr'+m_4(t',r')d\Omega\,,
\end{equation}
where  $m_i$ are functions of the radius $r'$ and of the time $t'$
(see also \cite{noether}). In GR, due to the freedom in the frame
choice, we can consider a coordinate transformation
$t\,=\,U_1(t',r')\,, \ r\,=\,U_2(t',r')$ which maps the metric
(\ref{me0}) in a new one where the off\,-\,diagonal term
$m_3(t',r')dt'dr'$ vanishes and $m_4(t',r')\,=\,-r^2$, that
is\footnote{This  condition allows to obtain the standard
definition of the circumference with  radius $r$.}\,:

\begin{equation}\label{me}
ds^2\,=\,A(t,r)dt^2-B(t,r)dr^2-r^2d\Omega\,.
\end{equation}
This  expression can be considered without loss of generality as
the most general definition of a spherically symmetric metric
compatible with a pseudo\,-\,Riemannian manifold  without torsion.
Actually, by inserting this metric into the field equations
(\ref{HOEQ}) and (\ref{TrHOEQ}), one obtains\,:
\begin{equation}\label{fe4}
f'(R)R_{\mu\nu}-\frac{1}{2}f(R)g_{\mu\nu}+\mathcal{H}_{\mu\nu}\,=\,\frac{\mathcal{X}}{2}T_{\mu\nu}\,,
\end{equation}

\begin{equation}\label{trfe4}
g^{\sigma\tau}H_{\sigma\tau}=f'(R)R-2f(R)+\mathcal{H}\,=\,\frac{\mathcal{X}}{2}T\,,
\end{equation}
where the two quantities $\mathcal{H}_{\mu\nu}$ and $\mathcal{H}$
read\,:
\begin{eqnarray}\label{highterms1}
\mathcal{H}_{\mu\nu}\,=\,-f''(R)\biggl\{R_{,\mu\nu}-\Gamma^t_{\mu\nu}R_{,t}-\Gamma^r_{\mu\nu}R_{,r}-
g_{\mu\nu}\biggl[\biggl({g^{tt}}_{,t}+g^{tt}
\ln\sqrt{-g}_{,t}\biggr)R_{,t}+\biggl({g^{rr}}_{,r}+g^{rr}\ln\sqrt{-g}_{,r}\biggr)R_{,r}+\nonumber\\\nonumber\\+g^{tt}R_{,tt}
+g^{rr}R_{,rr}\biggr]\biggr\}-f'''(R)\biggl[R_{,\mu}R_{,\nu}-g_{\mu\nu}\biggl(g^{tt}{R_{,t}}^2+g^{rr}
{R_{,r}}^2\biggr)\biggr]
\\\nonumber\\\label{highterms2}
\mathcal{H}\,=\,g^{\sigma\tau}\mathcal{H}_{\sigma\tau}\,=\,3f''(R)\biggl[\biggl({g^{tt}}_{,t}+g^{tt}
\ln\sqrt{-g}_{,t}\biggr)R_{,t}+\biggl({g^{rr}}_{,r}+g^{rr}\ln\sqrt{-g}_{,r}\biggr)R_{,r}+g^{tt}R_{,tt}
+g^{rr}R_{,rr}\biggr]+\nonumber\\\nonumber\\+
3f'''(R)\biggl[g^{tt}{R_{,t}}^2+g^{rr}{R_{,r}}^2\biggr]\,,
\end{eqnarray}
where $\Gamma^\alpha_{\mu\nu}$ are the standard Christoffel's
symbols related to the metric $g_{\mu\nu}$. These equations show
the interesting feature that the derivatives with respect to $R$
of $f(R)$ are very well distinct with respect to the time and
spatial derivatives of $R$. This peculiarity will allow to better
understand the dynamical behavior of solutions, as we shall see
below.

\section{The Ricci  curvature scalar in spherical symmetry}

As standard, the Ricci scalar can be written as

\begin{equation}\label{ricci}
R\,=\,g^{\mu\nu}R_{\mu\nu}\,=\,g^{\mu\nu}\left[\Gamma_{\mu\nu,\alpha}^{\alpha}-\Gamma_{\mu\alpha,\nu}^{\alpha}
+\Gamma^{\beta}_{\alpha\beta}\Gamma_{\mu\nu}^{\alpha}-\Gamma^{\alpha}_{\beta\mu}
\Gamma_{\nu\alpha}^{\beta}\right]\,,
\end{equation}
where  $\Gamma^\alpha_{\mu\nu}$ is the  Christoffel symbols
defined as
\begin{equation}\label{christhoffel}
\Gamma^\alpha_{\mu\nu}=\frac{1}{2}g^{\alpha\sigma}
\left(g_{\mu\sigma,\nu}+g_{\nu\sigma,\mu}-g_{\mu\nu,\sigma}\right)\;.
\end{equation}
Imposing the spherical symmetry (\ref{me}), the Ricci scalar in
terms of the gravitational potentials reads\,:

\begin{equation}\label{riccispher}
R(t,\,r)\,=\,\frac{B\biggl[\dot{A}\dot{B}-A'^2\biggr]r^2+A\biggl[r(\dot{B}^2-A'B')+2B(2A'+rA''-r\ddot{B})\biggr]
 -4A^2\biggl[B^2-B+r B'\biggr]}{2r^2A^2B^2}
\end{equation}
where, for sake of brevity, we have discarded the explicit
dependence in $A(t,\,r)$ and $B(t,\,r)$ and the prime indicates
the derivative with respect to $r$ while the dot with respect to
$t$. If the metric (\ref{me}) is time-independent, i.e.
$A(t,r)=a(r)$, $B(t,r)=b(r)$, Eq.(\ref{riccispher}) assumes the
simpler form\,:

\begin{equation}\label{ricscalin}R(r)=\frac{a(r)\biggl[2b(r)\biggl(2a'(r)+ra''(r)\biggr)-ra'(r)b'(r)\biggr]-b(r)a'(r)^2r^2
-4a(r)^2\biggl(b(r)^2-b(r)+rb'(r)\biggr)}{2r^2a(r)^2b(r)^2}
\end{equation}
where  the radial dependence of the gravitational potentials is
now explicitly shown. This expression can be seen as a constraint
for the functions $a(r)$ and $b(r)$ once a specific form of Ricci
scalar is given. In particular, it reduces to a Bernoulli equation
of index two, that is

\begin{displaymath}
b'(r)+h(r)b(r)+l(r)b(r)^2=0\,,
\end{displaymath}
with respect to the metric potential $b(r)$\,:

\begin{eqnarray}\label{eqric}
b'(r)+\biggl\{\frac{r^2a'(r)^2-4a(r)^2-2ra(r)[2a(r)'+ra(r)'']}{ra(r)[4a(r)
+ra'(r)]}\biggr\}b(r)+\biggl\{\frac{2a(r)}{r}\biggl[\frac{2+r^2R(r)}{4a(r)+ra'(r)}\biggr]\biggr\}b(r)^2\,=\,0\,.
\end{eqnarray}
A general solution of (\ref{eqric}) is:

\begin{equation}\label{gensol}
b(r)\,=\,\frac{\exp[-\int dr\,h(r)]}{K+\int dr\,l(r)\,\exp[-\int
dr\,h(r)]}\,,
\end{equation}
where $K$ is an integration constant while $h(r)$ and $l(r)$ are
the two functions which, respectively, define the coefficients of
the quadratic and the linear term with respect to $b(r)$, as in
the standard definition of the Bernoulli equation
\cite{bernoulli}. Looking at the equation, we can notice that it
is possible to have $l(r)\,=\,0$ which implies to find out
solutions with a Ricci scalar scaling as ${\displaystyle
-\frac{2}{r^2}}$ in term of the radial coordinate. On the other
side, it is not possible to have $h(r)\,=\,0$ since otherwise we
will get imaginary solutions. A particular consideration deserves
the limit $r\rightarrow\infty$. In order to achieve a
gravitational potential $b(r)$ with the correct Minkowski limit,
both $h(r)$ and $l(r)$ have to go to zero provided that the
quantity $r^2R(r)$ turns out to be constant: this result implies
$b'(r)=0$, and, finally, also the metric potential $b(r)$ has  a
correct Minkowski limit.

In general,  if we ask for the asymptotic flatness of the metric
as a feature of the theory,  the Ricci scalar has to evolve at
infinity as $r^{-n}$ with $n\geqslant 2$. Formally it has to be:

\begin{equation}\label{condricc}
\lim_{r\rightarrow\infty}r^2R(r)\,=\,r^{-n}
\end{equation}
with $n\in\mathbb{N}$. Any other behavior of the Ricci scalar
could affect the requirement of the correct asymptotic flatness.
This result can be easily deduced from Eq.(\ref{eqric}). In fact,
let us consider  the simplest spherically symmetric case:

\begin{equation}\label
{me3}ds^2\,=\,a(r)dt^2-\frac{dr^2}{a(r)}-r^2d\Omega\,.
\end{equation}
The Bernoulli Eq.(\ref{eqric}) is easy to integrate and the most
general metric potential $a(r)$, compatible with the Ricci scalar
constraint (\ref{ricscalin}), is\,:

\begin{equation}
a(r)\,=\,1+\frac{k_1}{r}+\frac{k_2}{r^2}+\frac{1}{r^2}\int \biggl[\int r^2 R(r)dr\biggr]dr
\end{equation}
where $k_1$ and $k_2$ are integration constants. Actually one gets
the standard result $a(r)=1$ (Minkowski) for $r\rightarrow \infty$
only if the condition (\ref{condricc}) is satisfied, otherwise we
get a diverging gravitational potential.

\section{Solutions with constant  curvature scalar}
The case of constant curvature is equivalent to GR with a
cosmological constant and the solution is time independent. This
result is well known (see, for example, \cite{barrottew}) but we
report, for the sake of completeness, some considerations related
with it in order to deal with more general case where a radial
dependence is supposed.  Let us assume a scalar curvature constant
($R\,=\,R_0$). The field Eqs.(\ref{fe4}) and (\ref{trfe4}), being
$\mathcal{H}_{\mu\nu}\,=\,0$,  reduce to:
\begin{equation}\label{fe2}
f'_0R_{\mu\nu}-\frac{1}{2}f_0g_{\mu\nu}\,=\,\frac{\mathcal{X}}{2}T_{\mu\nu}\,,
\end{equation}
\begin{equation}\label{fe2bis}
f'_0R_0-2f_0\,=\,\frac{\mathcal{X}}{2}T\,.
\end{equation}
where $f(R_0)=f_0$, $f'(R_0)=f'_0$. Such equations can be arranged
as:

\begin{equation}\label{fe3}
R_{\mu\nu}+\lambda g_{\mu\nu}=q\frac{\mathcal{X}}{2}T_{\mu\nu}
\end{equation}
\begin{equation}\label{fe3tr}R_0=q\frac{\mathcal{X}}{2}T-4\lambda
\end{equation}
where ${\displaystyle \lambda=-\frac{f_0}{2f'_0}}$ and
$q^{-1}=f'_0$. As standard, the stress-energy tensor of
perfect-fluid matter is
\begin{equation}\label{enmomten}
T_{\mu\nu}\,=\,(\rho+p)u_\mu u_\nu-pg_{\mu\nu}\,;
\end{equation}
where $\rho$ is the energy density, $ p$ is the pressure and
$u^\mu=dx^{\mu}/ds$ is the 4-velocity. A general solution of the
above set of equations is achieved for $p\,=\,-\rho$ and reads\,:

\begin{equation}
ds^2\,=\,\biggl(1+\frac{k_1}{r}+\frac{q\mathcal{X}\rho-2\lambda}{6}r^2\biggr)dt^2-\frac{dr^2}{1+\frac{k_1}{r}+
\frac{q\mathcal{X}\rho-2\lambda}{6}r^2}-r^2d\Omega\,.
\end{equation}
This result means that any $f(R)$\,-\,theory in the case of
constant curvature scalar ($R=R_0$) exhibits solutions with
cosmological constant\,-\,like behavior (de Sitter). This is one
of the reason why the dark energy issue can be addressed using
these theory \cite{f(R)-noi,f(R)-cosmo}. This fact is well known
using the FRW metric \cite{barrottew}. As another remark we have
to say that this solution cannot describe stellar structures since
this kind of equation of state does not work for stars so it could
be interesting only in cosmological contexts.

If $f(R)$ is analytic, it is:
\begin{equation}\label{f}
f(R)\,=\,\Lambda+\Psi_0R+\Psi(R)\,.
\end{equation}
where $\Psi_0$ is a coupling constant, $\Lambda$ plays the role of
the cosmological constant and $\Psi(R)$ is a generic analytic
function of $R$ satisfying the condition

\begin{equation}\label{psi}
\lim_{R\rightarrow 0}R^{-2}\Psi(R)\,=\,\Psi_1
\end{equation}
where $\Psi_1$ is a constant. If we neglect the cosmological
constant $\Lambda$ and $\Psi_0$ is set to zero, we obtain a new
class of theories which, in the limit $R\rightarrow{0}$, does not
reproduce GR (from  Eq.(\ref{psi}), we have $\lim_{R\rightarrow 0}
f(R)\sim R^2$). In such a case analyzing the whole set of
Eqs.(\ref{fe2}) and (\ref{fe2bis}), one can observe that both zero
and constant $\neq 0$ curvature solutions are possible. In
particular,  if $R\,=\,R_0\,=\,0$ field equations are solved for
every form of  gravitational potentials entering the spherically
symmetric  background,  provided that the Bernoulli equation
(\ref{eqric}), relating these functions, is fulfilled for the
particular case $R(r)=0$. The  solutions are thus defined by the
relation

\begin{equation}\label{gensol0}
b(r)\,=\,\frac{\exp[-\int
dr\,h(r)]}{K+4\int\frac{dr\,a(r)\,\exp[-\int
dr\,h(r)]}{r[a(r)+ra'(r)]}}\,,
\end{equation}
being $B(t,r)=b(r)$ from Eq.(\ref{me}). In Table \ref{tableR0}, we
give some examples of $f(R)$-theories admitting solutions with
constant$\neq 0$ or null scalar curvature. Each model admits
Schwarzschild, Schwarzschild\,-\,de Sitter, and the class of
solutions given by (\ref{gensol0}).
\begin{table}
\begin{center}
\begin{tabular}{|ccccc|}
  \hline
  & & & & \\
  & $f(R)$\,-\,theory: & & Field equations: & \\
  & & & & \\
  & $R$ & $\longrightarrow$ & $R_{\mu\nu}=0$, $\text{with}$ $R=0$ & \\
  & & & & \\
  & $\xi_1 R+\xi_2 R^n$ & $\longrightarrow$ & $\begin{cases}
                R_{\mu\nu}=0 & \text{with}\,\,R=0,\,\,\,\xi_1\neq 0\\
                R_{\mu\nu}+\lambda g_{\mu\nu}=0 &
                \text{with}\,\,R=\biggl[\frac{\xi_1}{(n-2)\xi_2}\biggr]^{\frac{1}{n-1}}
                ,\,\,\,\xi_1\neq 0,\,\,\,n\neq2\\
                0=0 & \text{with}\,\,R=0,\,\,\,\xi_1=0\\
                R_{\mu\nu}+\lambda g_{\mu\nu}=0 & \text{with}\ R=R_0,\ \xi_1=0,\ n=2
                \end{cases}$& \\
  & & & & \\
  & $\xi_1R+\xi_2R^{-m}$ & $\longrightarrow$ & $R_{\mu\nu}+\lambda g_{\mu\nu}=0$ with $R=\biggl[-\frac{(m+2)\xi_2}{\xi_1}\biggr]^{\frac{1}{m+1}}$ & \\
  & & & & \\
  & $\xi_1 R+\xi_2 R^n+\xi_3 R^{-m}$ & $\longrightarrow$ & $R_{\mu\nu}+\lambda g_{\mu\nu}=0$, \text{with}\ $R=R_0$\ so that\ $\xi_1R_0^{m+1}+(2-n)\xi_2R_0^{n+m}
  +(m+2)\xi_3=0$ & \\
  & & & & \\
  & $\frac{R}{\xi_1+R}$ & $\longrightarrow$ & $\begin{cases}
                                         R_{\mu\nu}=0 & \text{with}\ R=0\\
                                         R_{\mu\nu}+\lambda g_{\mu\nu}=0 & \text{with}\ R=-\frac{\xi_1}{2}
                                         \end{cases}$ & \\
  & & & & \\
  & $\frac{1}{\xi_1+R}$ & $\longrightarrow$ & $R_{\mu\nu}+\lambda g_{\mu\nu}=0$, \text{with}\ $R=-\frac{2\xi_1}{3}$ & \\
  & & & & \\
  \hline
\end{tabular}
\end{center}
\caption{\label{tableR0} Emamples of $f(R)$-models admitting
constant and zero scalar curvature solutions. In the right hand
side,  the field equations are given for each model. The power
$n$, $m$ are natural numbers while $\xi_i$ are generic real
constants.}
\end{table}

\section{Solutions with  curvature scalar function of  $r$}

Up to now we have discussed the behavior of $f(R)$ gravity seeking
for spherically symmetric solutions with  constant scalar
curvature. This situation is well known in GR  and give rise to
the  Schwarzschild solution $(R=0)$ and the Schwarzschild\,-\,de
Sitter solution $(R=R_0\neq 0)$. The problem can be generalized in
$f(R)$ gravity  considering  the Ricci scalar as an arbitrary
function of the radial coordinate $r$.

This approach is  interesting since, in general, Higher Order
Gravity theories are supposed to admit such kind of solutions and
several examples have been found in  literature
\cite{ghost-f(R),noi-mnras,multamaki06,noether}. Here we want to
face the problem from general point of view.

If we choose the Ricci scalar $R$ as a generic function of the
radial coordinate ($R=R(r)$), it is possible to show that also in
this case the  solution of the field Eqs.(\ref{fe4}) and
(\ref{trfe4}) is time independent (if $T_{\mu\nu}=0$). In other
words,   the Birkhoff theorem has to hold. The crucial point of
the approach is to study the off\,-\,diagonal
$\{t,r\}$\,-\,component of (\ref{fe4}) as well as in the case of
GR. This equation, for a generic $f(R)$  reads\,:

\begin{equation}
\frac{d}{dr}\biggl(r^2f'(R)\biggr)\dot{B}(t,r)\,=\,0\,,
\end{equation}
and two possibilities are in order. Firstly, we can choose
$\dot{B}(t,r)\neq 0$. This choice implies that ${\displaystyle
f'(R)\sim\frac{1}{r^2}}$. If this is the case, the remaining field
equation turn out to be not fulfilled and it can be easily
recognized that the dynamical system encounters a mathematical
incompatibility.

The only possible solution is given by $\dot{B}(t,\,r)\,=\,0$ and
then the gravitational potential has to be $B(t,r)\,=\,b(r)$.
Considering also the $\{\theta,\theta\}$\,-\,equation one can
determine that the gravitational potential $A(t,r)$ can be
factorized with respect to the time, so that we get  solutions
which can be recast in the stationary spherically symmetric form
after a suitable coordinate transformation.

As a matter of fact, even the more general radial dependent case
admit  time\,-\,independent solutions. From the trace equation and
the $\{\theta,\theta\}$\,-\,component, we deduce a relation which
links $A(t,r)=a(r)$ and $B(t,r)=b(r)$\,:

\begin{equation}\label{birk}
a(r)\,=\,\frac{b(r)e^{\frac{2}{3}\int\frac{(Rf'-2f)b(r)}{R'f''}dr}}{r^4R'^2f''^2}\,,
\end{equation}
(with $f''\neq 0$) and one which relates $b(r)$ and $f(R)$ (see
also \cite{multamaki06} for a similar result)\,:

\begin{equation}\label{rebf}
b(r)\,=\,\frac{6[f'(rR'f'')'-rR'^2f''^2]}{rf(rR'f''-4f')+2f'(rR(f'-rR'f'')-3R'f'')}\,.
\end{equation}

As above,   three further equations has to be satisfied to
completely solve the system (respectively the $\{t,t\}$ and
$\{r,r\}$ components of the field equations and the Ricci scalar
constraint) while  the only unknown functions are $f(R)$ and the
Ricci scalar $R(r)$.

 If we now consider a fourth order model
of the form $f(R)\,=\,R+\Phi(R)$, with $\Phi(R)\ll R$ we are
capable of satisfying the whole  set of equations up to third
order in $\Phi$. In particular,  we can solve the whole set of
equations: the relations (\ref{birk}) and (\ref{rebf}) will give
the general solution  depending only on the forms of $\Phi(R)$ and
$R\,=\,R(r)$, that is:

\begin{eqnarray}\label{solRr}
{}a(r)\,&=&\,\displaystyle\frac{b(r)e^{\displaystyle-\frac{2}{3}\int\frac{[R+(2\Phi-R\Phi')]b(r)}{R'\Phi''}dr}}{r^4R'^2\Phi''^2}
\\\nonumber\\
b(r)\,&=&\,-\displaystyle\frac{3 (rR'\Phi'')_{,r}}{rR}\,.
\end{eqnarray}
This solution is one of the main results of this paper. In fact,
once the radial dependence of the scalar curvature is obtained,
Eqs.(\ref{solRr}) allow to write down the solution of the field
equations  and  the  gravitational potential, related to the
function $a(r)$, can be deduced. Furthermore one can check the
physical relevance of such a potential by means of astrophysical
data, see for example the analysis in \cite{kainulainen}. As a
final remark, we have to stress that the solution (\ref{solRr})
has been achieved here in the vacuum case but, in general, $R(r)$
should be obtained from the source $T$ using the trace
Eq.(\ref{TrHOEQ}). This task is simpler in Palatini approach than
in the metric approach since, in the former case, the box operator
is not present in the trace equation. For a detailed discussion,
see \cite{spher-f(R),bustelo}.

\section{Perturbing the spherically symmetric solutions}

The search for solutions in $f(R)$-gravity, in the case of Ricci
scalar dependent on the radial coordinate, can be faced by means
of a perturbation approach.   There are several perturbation
techniques by which higher order gravity  can be investigated in
the weak field limit. A general approach is starting from
analytical  $f(R)$ theories assuming that the background model
slightly deviates from the Einstein GR  (this means to consider
$f(R)= R+\Phi(R)$ where $\Phi(R)\ll R$ as above). Another approach
can be developed starting from the background metric considered as
the 0th\,-\,order solution. Both these approaches assume the weak
field limit of a given higher order gravity theory as a correction
to the Einstein GR, supposing that zero order approximation should
yield the standard lore.

Both these methods can provide interesting results on the
astrophysical scales where spherically symmetric solutions
characterized by small values of the scalar curvature, can be
taken into account.

In the following, we will consider the first  approach assuming
that the background metric matches, at zero order, the GR
solutions.

In general, searching for  solutions by a perturbation technique
means to perturb the metric
$g_{\mu\nu}=g^{(0)}_{\mu\nu}+g^{(1)}_{\mu\nu}$. This implies that
the field equations (\ref{fe4}) and (\ref{trfe4}) split, up to
first order,  in two levels. Besides, a perturbation on the metric
acts also on the Ricci scalar $R$ (see the relation (\ref{ricci}))
and then we can Taylor expand the analytic $f(R)$ about the
background value of $R$, i.e.:
\begin{equation}\label{approx}
f(R)=\sum_{n}\frac{f^{n(0)}}{n!}\biggl[R-R^{(0)}\biggr]^n=\sum_{n}\frac{f^{n(0)}}{n!}{R^{(1)}}^n\,.
\end{equation}
However the above condition $\Phi(R)\ll R$ has to imply the
validity of the linear approximation
$f''(R^{(0)})/f'(R^{(0)})R^{(1)}\ll 1$. This is demonstrated by
assuming $f'(R)=1+\Phi'(R)$ and $f''(R)=\Phi''(R)$. Immediately we
obtain that the condition is fulfilled for
\begin{equation}\label{condition}
\frac{\Phi''(R^{(0)})R^{(1)}}{1+\Phi'(R^{(0)})}\ll 1\,.
\end{equation}
For example, given a Lagrangian of the form ${\displaystyle
f(R)=R+\frac{\mu}{R}}$, it means
\begin{equation}
\frac{2\mu R^{(1)}}{R^{(0)}(R^{(0)^2}-\mu)}\ll 1\,,
\end{equation}
while, for $f(R)=R+\alpha R^2$, it is
\begin{equation}
\frac{2\alpha R^{(1)}}{1+2\alpha R^{(0)}}\ll 1\,.
\end{equation}
This means that  the validity of the approximation strictly
depends on the form of the models and the value of the parameters,
in the previous case $\mu$ and $\alpha$. For the considerations
below, we will assume that it holds. A detailed discussion for the
Palatini formalism is in \cite{spher-f(R),bustelo}.

The zero order field equations read\,:

\begin{equation}\label{eqp0}
f'^{(0)}R^{(0)}_{\mu\nu}-\frac{1}{2}g^{(0)}_{\mu\nu}f^{(0)}+\mathcal{H}^{(0)}_{\mu\nu}\,=\,\frac{\mathcal{X}}{2}T^{(0)}_{\mu\nu}
\end{equation}
where

\begin{eqnarray}
\mathcal{H}^{(0)}_{\mu\nu}\,=\,-f''^{(0)}\biggl\{R^{(0)}_{,\mu\nu}-{\Gamma^{(0)}}^{\rho}_{\mu\nu}R^{(0)}_{,\rho}-g^{(0)}_{\mu\nu}
\biggl({g^{(0)\rho\sigma}}_{,\rho}R^{(0)}_{,\sigma}+g^{(0)\rho\sigma}R^{(0)}_{,\rho\sigma}+g^{(0)\rho\sigma}\ln\sqrt{-g}_{,\rho}R^{(0)}
_{,\sigma}\biggl)\biggl\}+\nonumber\\\nonumber\\-f'''^{(0)}\biggl\{R^{(0)}_{,\mu}R^{(0)}_{,\nu}-g^{(0)}_{\mu\nu}g^{(0)\rho\sigma}R^{(0)}_{,\rho}R^{(0)}
_{,\sigma} \biggl\}\,.
\end{eqnarray}
At first order one has\,:

\begin{equation}\label{eqp1}
f'^{(0)}\biggl\{R^{(1)}_{\mu\nu}-\frac{1}{2}g^{(0)}_{\mu\nu}R^{(1)}\biggr\}+f''^{(0)}R^{(1)}R^{(0)}_{\mu\nu}
-\frac{1}{2}f^{(0)}g^{(1)}_{\mu\nu}+\mathcal{H}^{(1)}_{\mu\nu}\,=\,\frac{\mathcal{X}}{2}T^{(1)}_{\mu\nu}
\end{equation}
with

\begin{eqnarray}
\mathcal{H}^{(1)}_{\mu\nu}\,=\,-f''^{(0)}\biggl\{R^{(1)}_{,\mu\nu}-{\Gamma^{(0)}}^{\rho}_{\mu\nu}R^{(1)}_{,\rho}-{\Gamma^{(1)}}^{\rho}
_{\mu\nu}R^{(0)}_{,\rho}-g^{(0)}_{\mu\nu}\biggl[{g^{(0)\rho\sigma}}_{,\rho}R^{(1)}_{,\sigma}+{g^{(1)\rho\sigma}}_{,\rho}R^{(0)}_{,\sigma}+g^{(0)\rho\sigma}
R^{(1)}_{,\rho\sigma}+\nonumber\\\nonumber\\+g^{(1)\rho\sigma}R^{(0)}_{,\rho\sigma}+g^{(0)\rho\sigma}\biggl(\ln\sqrt{-g}^{(0)}_{,\rho}R^{(1)}_{,\sigma}+
\ln\sqrt{-g}^{(1)}_{,\rho}R^{(0)}_{,\sigma}\biggl)+g^{(1)\rho\sigma}\ln\sqrt{-g}^{(0)}_{,\rho}R^{(0)}_{,\sigma}\biggr]-g^{(1)}_{\mu\nu}\biggl({g^{(0)\rho
\sigma}}_{,\rho}R^{(0)}_{,\sigma}+\nonumber\\\nonumber\\+g^{(0)\rho\sigma}R^{(0)}_{,\rho\sigma}+g^{(0)\rho\sigma}\ln\sqrt{-g}^{(0)}_{,\rho}R^{(0)}_
{,\sigma}\biggr)\biggr\}-f'''^{(0)}\biggl\{R^{(0)}_{,\mu}R^{(1)}_{,\nu}+R^{(1)}_{,\mu}R^{(0)}_{,\nu}-g^{(0)}_{\mu\nu}g^{(0)\rho\sigma}\biggl(R^{(0)}_
{,\rho}R^{(1)}_{,\sigma}+R^{(1)}_{,\rho}R^{(0)}_{,\sigma}\biggr)+\nonumber\\\nonumber\\-g^{(0)}_{\mu\nu}g^{(1)\rho\sigma}R^{(0)}_{,\rho}R^{(0)}_{,\sigma}
-g^{(1)}_{\mu\nu}g^{(0)\rho\sigma}R^{(0)}_{,\rho}R^{(0)}_{,\sigma}\biggr\}-f'''^{(0)}R^{(1)}\biggl\{R^{(0)}_{,\mu\nu}-{\Gamma^{(0)}}^{\rho}_{\mu\nu}R^{(0)}
_{,\rho}-g^{(0)}_{\mu\nu}\biggl({g^{(0)\rho\sigma}}_{,\rho}R^{(0)}_{,\sigma}+\nonumber\\\nonumber\\+g^{(0)\rho\sigma}R^{(0)}_{,\rho\sigma}+g^{(0)\rho
\sigma}\ln\sqrt{-g}_{,\rho}R^{(0)}_{,\sigma}\biggl)\biggl\}-f^{IV(0)}R^{(1)}\biggl\{R^{(0)}_{,\mu}R^{(0)}_{,\nu}-g^{(0)}_{\mu\nu}g^{(0)\rho\sigma}R^{(0)}_
{,\rho}R^{(0)}_{,\sigma}\biggl\}\,.
\end{eqnarray}
A part the analyticity, no  hypothesis has been invoked on the
form of $f(R)$.  As a matter of fact, $f(R)$ can be completely
general. At this level, to solve the problem, it is required  the
zero order solution of Eqs.(\ref{eqp0}) which, in general, could
be a GR solution. This problem can be overcome assuming the same
order  of perturbation on the $f(R)$, that is:

\begin{equation}\label{theapprox}
f\,=\,R+\Phi(R)\,,
\end{equation}
where $\Phi(R)$ is a generic function of the Ricci scalar
fulfilling the prescription $\Phi\ll R$. Then we have
\begin{equation}\label{approx2}
f\,=\,R^{(0)}+R^{(1)}+\Phi^{(0)}\,,\ \ \ f'\,=\,1+\Phi'^{(0)}\,,\
\ \ f''\,=\,\Phi''^{(0)}\,,\ \ \ \ \ \ f'''=\Phi'''^{(0)}\,,
\end{equation}
and Eqs.(\ref{eqp0})  reduce to the form

\begin{equation}
R^{(0)}_{\mu\nu}-\frac{1}{2}R^{(0)}g^{(0)}_{\mu\nu}\,=\,\frac{\mathcal{X}}{2}T^{(0)}_{\mu\nu}\,.
\end{equation}
On the other hand,  Eqs.(\ref{eqp1}) reduce to

\begin{equation}\label{}
R^{(1)}_{\mu\nu}-\frac{1}{2}g^{(0)}_{\mu\nu}R^{(1)}-\frac{1}{2}g^{(1)}_{\mu\nu}R^{(0)}-\frac{1}{2}g^{(0)}_{\mu\nu}
\Phi^{(0)}+\Phi'^{(0)}R^{(0)}_{\mu\nu}+\mathcal{H}^{(1)}_{\mu\nu}\,=\,\frac{\mathcal{X}}{2}T^{(1)}_{\mu\nu}
\end{equation}
where

\begin{eqnarray}
\mathcal{H}^{(1)}_{\mu\nu}\,=\,-\Phi'''^{(0)}\biggl\{R^{(0)}_{,\mu}R^{(0)}_{,\nu}-g^{(0)}_{\mu\nu}g^{(0)rr}R^{(0)}_{,r}R^{(0)}_{,r}
\biggr\}-\Phi''^{(0)}\biggl\{R^{(0)}_{,\mu\nu}-\Gamma^{(0)r}_{\mu\nu}R^{(0)}_{,r}-g^{(0)}_{\mu\nu}\biggl({g^{(0)}}^{rr}_{,r}R^{(0)}_{,r}+
\nonumber\\\nonumber\\+g^{(0)rr}R^{(0)}_{,rr}+g^{(0)rr}\ln\sqrt{-g}^{(0)}_{,r}R^{(0)}_{,r}\biggl)\biggr\}\,.
\end{eqnarray}
The new system of field equations is evidently simpler than the
starting one and once  the zero order solution is obtained, the
solutions at the first order correction can be easily achieved. In
Table \ref{table2},  a list of  solutions, obtained with this
perturbation method, is given considering different classes of
$f(R)$ models.

Some remarks on these solutions are in order at this point. In the
case of $f(R)$ models which  are evidently corrections to the
Hilbert-Einstein Lagrangian as $\Lambda+R+\epsilon R \ln R$ and
$R+\epsilon R^n$, with $\epsilon\ll 1$, one obtains  exact
solutions for the gravitational potentials $a(r)$ and $b(r)$
related by $a(r)=b(r)^{-1}$. The first order expansion is
straightforward as in GR.  If the functions $a(r)$ and $b(r)$ are
not related, for $f(R)\,=\,\Lambda+R+\epsilon R \ln R$, the first
order system is directly solved  without any prescription on the
perturbation functions $x(r)$ and $y(r)$. This is not the case for
$f(R)\,=\,R+\epsilon R^n$ since, for this model, one can obtain an
explicit constraint on the perturbation function  implying the
possibility to deduce the form of the gravitational potential
$\phi(r)$ from ${\displaystyle a(r)\,=\,1+\frac{2\phi(r)}{c^2}}$.
In such a case, no corrections are found with respect to the
standard Newtonian potential. The  theories $f(R)\,=\,R^n$ and
${\displaystyle f(R)\,=\,\frac{R}{(\xi+R)}}$ show similar
behaviors. The case $f(R)=R^2$ is peculiar and it has to be dealt
independently.

\begin{table}
\begin{center}
\begin{tabular}{|cccc|}
  \hline
  & & & \\
  & $f(R)$ - theory: & $\Lambda+R+\epsilon R \ln R$ & \\
  & & & \\
  & spherical potentials: & $a(r)=b(r)^{-1}=1+\frac{k_1}{r}-\frac{\Lambda r^2}{6}+\delta x(r)$ & \\
  & & & \\
  & solutions: & $x(r)=\frac{k_2}{r}+\frac{\epsilon\Lambda[\ln(-2\Lambda)-1]r^2}{6\delta}$ & \\
  & & & \\
  & first order metric: & $a(r)=1-\frac{\Lambda r^2}{6}+\delta x(r)$,\,\,\,\,\,$b(r)=\frac{1}{1-\frac{\Lambda r^2}{6}}+\delta y(r)$ & \\
  & & & \\
  & solutions: & $\begin{cases} x(r)=(\Lambda r^2-6)\biggl\{k_1+\int dr\frac{4\delta(2\Lambda^2r^4-15\Lambda r^2+18)y(r)+r\{36r\epsilon\Lambda[\log(-
  2\Lambda)-1]+\delta(\Lambda r^2-6)^2y'(r)\}}{36r\delta(\Lambda r^2-6)}\biggr\}\\
  y(r)=\frac{k_2\delta-6r^3\epsilon\Lambda[\ln(-2\Lambda)-1]}{r\delta(r^2\Lambda-6)^2}  \end{cases} $& \\
  & & & \\
  & & & \\
  & $f(R)$ - theory: & $R+\epsilon R^n$ & \\
  & & & \\
  & spherical potentials: & $a(r)=b(r)^{-1}=1+\frac{k_1}{r}+\delta x(r)$ & \\
  & & & \\
  & solutions: & $x(r)=\frac{k_2}{r}$ & \\
  & & & \\
  & first order metric: & $a(r)=1+\delta\frac{x(r)}{r}$,\,\,\,\,\,$b(r)=1+\delta\frac{y(r)}{r}$ & \\
  & & & \\
  & solutions: & $x(r)=k_1+k_2r$,\,\,\,\,\,$y(r)=k_3$ & \\
  & & & \\
  & & & \\
  & $f(R)$ - theory: & $R^n$ & \\
  & & & \\
  & spherical potentials: & $a(r)=b(r)^{-1}=1+\frac{k_1}{r}+\frac{R_0r^2}{12}+\delta x(r)$ & \\
  & & & \\
  & solutions: & $\begin{cases} n=2,\,\,\,\,\,\,R_0\neq 0 \,\,\text{and}\,\,x(r)=\frac{3k_2-k_3}{3r}+\frac{k_3r^2}{12}+\frac{k_4}{r}\int
  dr\,r^2\biggl\{\int
  dr\frac{\exp\biggl[\frac{R_0r_0^2\ln(r-r_0)}{8+3R_0r_0^2}\biggr]}{r^5}\biggr\}\\
  \,\,\,\,\,\,\,\,\,\,\,\,\,\,\,\,\,\,\,\,\,\,\,\, \text{with}\,\,r_0\,\,$satisfying the condition$\,\,6k_1+8r_0+R_0r_0^3=0 \\n\geq 2,\,\,\,\,\,\,
  \text{System solved only whit $R_0=0$ and no prescriptions on  $x(r)$}  \end{cases}$ & \\
  & & & \\
  & first order metric: & $a(r)=1+\delta\frac{x(r)}{r}$,\,\,\,\,\,$b(r)=1+\delta\frac{y(r)}{r}$ & \\
  & & & \\
  & solutions: & $\begin{cases} n=2 & y(r)=-\frac{R_0r^3}{6}-\frac{x(r)}{2}+\frac{1}{2}rx'(r)+k_1,\,\,R(r)=\delta R_0 \\ n\neq 2 &
  y(r)=-\frac{1}{2}\int dr\,r^2R(r)-\frac{x(r)}{2}+\frac{1}{2}rx'(r)+k_1\,\,\,\text{with $R(r)$ whatever}\end{cases}$ & \\
  & & & \\
  & first order metric: & $a(r)=1-\frac{r_g}{r}+\delta x(r)$,\,\,\,\,\,$b(r)=\frac{1}{1-\frac{r_g}{r}}+\delta y(r)$ & \\
  & & & \\
  & solutions: & $\begin{cases} n=2 & y(r)=\frac{rk_1}{3r_g^2-7r_gr+4r^2}+\frac{r^2k_2}{3(3r_g^2-7r_gr+4r^2)}+\frac{r_gr^2x(r)+2(r_gr^3-r^4)x'(r)}{(3r_g-4r)
  (r_g-r)^2} \\ n\neq 2 &
  \text{whatever functions $x(r)$\,, $y(r)$ and $R(r)$} \end{cases}$ & \\
  & & & \\
  & & & \\
  & $f(R)$ - theory: & $R/(\xi+R)$ & \\
  & & & \\
  & first order metric: & $a(r)=1+\delta\frac{x(r)}{r}$,\,\,\,\,\,$b(r)=1+\delta\frac{y(r)}{r}$ & \\
  & & & \\
  & solutions: & $\begin{cases} x(r)=-\frac{4e^{-\frac{\xi^{1/2}r}{\sqrt{6}}}}{\xi}k_1-\frac{2\sqrt{6}e^{\frac{\xi^{1/2}r}{\sqrt{6}}}}{\xi^{3/2}}k_2+k_3r \\
  y(r)=-\frac{2e^{-\frac{\xi^{1/2}r}{\sqrt{6}}}(6\xi^{1/2}+\sqrt{6}\xi\,r)}{3b^{3/2}}k_1-\frac{2e^{\frac{\xi^{1/2}r}{\sqrt{6}}}(\sqrt{6}-\xi^{1/2}r)}
  {\xi^{3/2}}k_2
  \end{cases}$ & \\
  & & & \\
  \hline
\end{tabular}
\end{center}
\caption{A list of exact solutions obtained {\it via} the
perturbation approach for several classes of $f(R)$ theories;
$k_i$ are integration constants and $r_g=\frac{2GM}{c^2}$ is the
Schwarzschild radius; the potentials $A(t,r)=a(r)$ and
$B(t,r)=b(r)$ are defined by the metric
(\ref{me}).}\label{table2}\end{table}

\newpage

\section{Conclusions}

General Relativity has been consistently tested in physical
situations implying, essentially, spherical symmetry and weak
field limit \cite{will}. One of the fundamental and obvious issue
that any theory of gravity should satisfy is the fact that, in
absence of gravitational field or very far from a given
distribution of sources, the spacetime has to be asymptotically
flat (Minkowski). Any alternative or modified gravitational theory
(beside the diffeomorphism invariance and the general covariance)
should address these physical requirements to be consistently
compared with GR. This is a crucial point which several times is
not considered when people is constructing the weak field limit of
 alternative theories of gravity.

In this paper, we have faced this problem discussing, in the most
general way, what  the meaning of spherical solutions is in $f(R)$
theories of gravity and when the standard results of GR are
recovered in the limits $r\rightarrow\infty$ and $f(R)\rightarrow
R$.  Essentially, spherical solutions can be classified, with
respect to the Ricci curvature scalar $R$, as $R=0$, $R=R_0\neq
0$, and $R=R(r)$, where $R_0$ is a constant and $R(r)$ is a
function of the radial coordinate $r$. In these cases, the
Birkhoff theorem holds; this means that stationary solutions are
also static. However, as shown in the companion paper
\cite{Newf(R)}, this theorem does not hold, in general, for every
$f(R)$ theory since time-dependent evolution can emerge depending
on the order of perturbations.

In order to achieve exact spherical solutions, a crucial role is
played by the relations existing between the metric potentials and
between them and the Ricci curvature scalar. In particular, the
relations between the metric potentials and the Ricci scalar can
be used as a constraint: this gives a Bernoulli equation. Solving
it, in principle,  spherically symmetric solutions can be obtained
for any analytic $f(R)$ function, both for constant curvature
scalar and for curvature scalar depending on $r$.

Such spherically symmetric solutions can be used as background to
test how generic $f(R)$ theories of gravity deviate from GR.
Particularly interesting are  those theories that imply
$f(R)\rightarrow R$ in the weak field limit. In such cases, the
experimental comparison is straightforward and also experimental
results, evading GR constraints, can be framed in a
self-consistent picture \cite{bertolami}.

Finally, we have constructed a perturbation approach in which we
search for spherically symmetric solutions at the 0th-order and
then we search for solutions at the first order. The scheme is
iterative and could be, in principle, extended to any order in
perturbations. The crucial request is to take into account $f(R)$
theories which are Taylor expandable about some value $R=R_0$ of
the curvature scalar.

A important remark is in order at this point. Considering interior
and exterior solutions, the junction conditions are related to the
integration constants of the problem and strictly depend on the
source (e.g. the form of $T$). We have not considered this aspect
here since we have, essentially, searched for vacuum solutions.
However, such a problem has to be carefully faced in order to deal
with physically consistent solutions. For example, the
Schwarzschild solution $R=0$, which is one of the exterior
solutions which we have considered, always satisfies the junction
conditions with physically interesting interior metric. This is
not the case for several spherically symmetric solutions which
could give rise to unphysical junction conditions and not be in
agreement with Newton's law of gravitation, also asymptotically.
In these cases, such solutions have to be discarded. A detailed
analysis in this sense is in the paper \cite{barraco} for
spherically symmetric solutions obtained in the Palatini formalism
and in \cite{havas} for the metric approach.

\end{document}